\documentclass[twoside,a4paper,11pt]{proceedings}
\usepackage{graphicx}
\usepackage{hyperref}
\usepackage{movie15}
\usepackage{natbib}

\usepackage[font=small,labelfont=bf]{caption} % Required for specifying captions to tables and figures
\usepackage{booktabs} % Horizontal rules in tables
\usepackage{relsize} % Used for making text smaller in some places

\usepackage{amssymb}
\usepackage{xcolor}

\newcommand{\pp}{$pp$\,}

\begin{document}
\pagenumbering{arabic}
\pagestyle{myheadings}
\thispagestyle{empty}
\vspace*{-1cm}
%{\flushleft\includegraphics[width=\textwidth,bb=58 650 590 680]{stamp.pdf}}
%{\flushleft\includegraphics[width=\textwidth,viewport=58 650 590 680]{stamp.pdf}}
{\flushleft\includegraphics[width=.55\textwidth]{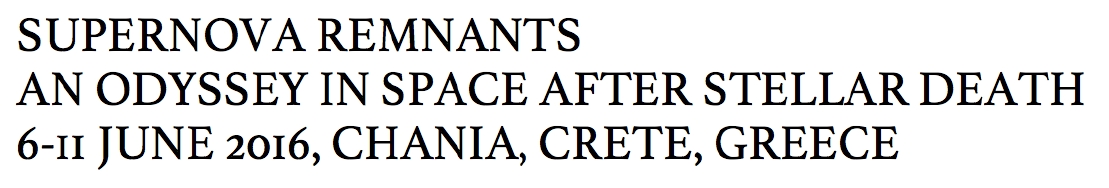}}
\vspace*{0.2cm}
\begin{flushleft}
{\bf {\LARGE
%%% TITLE of the paper. 
Early-time signatures of $\gamma$-ray emission from supernovae  in dense circumstellar media
}\\
\vspace*{1cm}
%%% Include here the LIST OF AUTHORS.
%%% Note that the last author has to be preceeded by an AND.
Dimitrios Kantzas$^{1\dag}$,
Maria Petropoulou$^{2\ast}$ and
Apostolos Mastichiadis$^1$
%and Tsikoudias$^4$
%,
%
% Do not delete next few lines
}\\
\vspace*{0.5cm}
%
%%% AFFILIATIONS LIST.
%%% and the AFFILIATIONS LIST. Note that one affiliation per line.
%%% Add as many affiliations as necessary. 
$^{1}$
Department of Physics, University of Athens, 15783 Zografos, Greece \\
$^2$
Department of Physics and Astronomy, Purdue University, West Lafayette, IN 47907, USA \\
%$^{3}$
%Very Expensive University, Euphoria State, USA\\
$^{\dag}$dimitrisk07@hotmail.com\\
$^{\ast}$mpetropo@purdue.edu
%
% Do not delete next few lines
\end{flushleft}
% Headings
\markboth{
%%% Type the SHORT version of the paper t
$\gamma$-rays from SNe with dense CSM
}{
%%%  First Author \& Second Author   OR   First-author et al. 
%%%  First Author \& Second Author   OR   First-author et al. if the author list contains three or more authors.
D. Kantzas et al.
}
\thispagestyle{empty}
\vspace*{0.4cm}
\begin{minipage}[l]{0.09\textwidth}
\ 
\end{minipage}
\begin{minipage}[r]{0.9\textwidth}
\vspace{1cm}
\section*{Abstract}{\small
%%% Type the ABSTRACT of your paper
We present our results on the $\gamma$-ray emission from interaction-powered supernovae (SNe), a recently discovered SN type that is suggested to be surrounded by a  circumstellar medium (CSM) with densities $10^7-10^{12}$~cm$^{-3}$. Such high densities favor inelastic collisions between relativistic protons accelerated in the SN blast wave and CSM protons and the production of $\gamma$-ray photons through neutral pion decays.  Using a numerical code that includes synchrotron radiation, adiabatic losses due to the expansion of the source, photon-photon  interactions, proton-proton collisions and proton-photon interactions, we calculate the multi-wavelength non-thermal photon emission soon after the shock breakout and follow its temporal evolution until 100-1000 days. Focusing on the $\gamma$-ray emission at $>100$ MeV, we show that this could be detectable by the {\it Fermi}-LAT telescope for nearby ($\lesssim 10$ Mpc) SNe with dense CSM ($>10^{11}$~cm$^{-3}$).

\vspace{10mm}
\normalsize}
\end{minipage}
%%% BODY of the paper
\section{Introduction}
\vspace{-0.1in}
The interaction of the shock waves of young supernovae (SNe) with the circumstellar medium (CSM) is the main driving power of the observed thermal and non-thermal emission that is detected  predominantly in radio and X-rays. The shocked CSM is heated to high temperatures ($\sim 10^7$~K), while the shock-accelerated electrons ({\it primaries}) emit synchrotron radiation that typically peaks in the radio bands. Together with electrons,  protons  of the CSM are also accelerated at the shock to multi-TeV/PeV energies (for a review see \cite{blasi2013}). Inelastic proton-proton  (\pp) collisions between the accelerated and non-relativistic protons of the shocked CSM  lead to the production of pions ($\pi^{+},\pi^{-},\pi^{0}$), which further decay into lighter particles including  $e^{-} e^{+}$ pairs ({\it secondaries}), neutrinos and $\gamma$-rays. Due to the small cross section for \pp  collisions ($\sigma_{pp}\simeq 3\times 10^{-26}$~cm$^2$), high CSM densities are required for a non-negligible contribution of secondary 
particles to the observed emission.  Such conditions may be realized whenever a shock plunges through a dense stellar wind. As SNe occur in CSM with a wide range of densities, spreading over at least six orders of magnitude, they provide a promising testing ground for cosmic-ray (CR) acceleration theories through the detection of multi-messenger signatures produced by the secondary particles.  

Although the smoking gun for CR acceleration in interaction-powered SNe would be the detection of high-energy neutrinos, a firm association of the IceCube events with one (or more) astrophysical candidate classes of sources is still lacking \citep[e.g.][]{aartsen2015a, aartsen2015b}. Yet, the radio synchrotron emission from secondary $e^{-}$ and the $\gamma$-ray emission from $\pi^0$ decays may be observable under certain conditions. Our preliminary semi-analytic estimates of the former are encouraging, as we have predicted observable features in the radio-mm spectra and light curves \citep{petropoulou2016radio}. 

Here, we present the first results of a theoretical investigation of the $\gamma$-ray emission in the energy band of \emph{Fermi}-LAT. 
For this purpose, we have been developing an one-zone model for the time-dependent calculation of the multi-wavelength (MW) non-thermal emission in SNe with dense CSM, which makes testable predictions in the \emph{Fermi}-LAT band. Ultimately, the model predictions can be used to probe CR acceleration in young SNe with dense CSM and, in particular, to indirectly estimate microphysical parameters, such as the (primary) electron-to-proton ratio $K_{ep}$ and the CR acceleration efficiency $\epsilon_p$.

\section{Theoretical framework}
\vspace{-0.1in}
The interaction of the freely-expanding SN ejecta with the CSM (modelled as an extended shell of matter) gives rise to two shock waves, i.e. a fast shock wave in the circumstellar material (forward shock) and a  reverse shock in the outer parts of the SN ejecta.  As long as the interaction between the SN ejecta and the CSM takes place within a region that is optically thick to Thomson scattering, the SN shock will be radiation-mediated and particle acceleration will be suppressed \citep[e.g.][]{murase11, katz11}. For dense CSM environments the shock is expected to breakout in the progenitor wind \citep{weaver76}; the previously radiation-mediated shock becomes collisionless  and particle acceleration can, in principle, take place.  

We assume that a fraction of the incoming particles (both electrons and protons) at the SN forward shock will be accelerated into a power-law distribution in energy (e.g. by the first order Fermi process \cite[e.g.][]{axford77, bell78}). In addition, the accelerated protons acquire a fraction $\epsilon_p$ of the incoming shock kinetic energy, while the shock-accelerated electron-to-proton ratio is $K_{ep}=10^{-2}K_{ep,-2}$. The shock-accelerated protons are advected in the downstream region of the shock where they can interact with the thermal (non-relativistic) protons of the shocked  CSM producing energetic pions. These, in return, decay into other secondary particles: $\pi^{\pm} \rightarrow \mu^{\pm} + \nu_{\mu} (\bar{\nu_\mu})$, $\mu^{\pm} \rightarrow e^{\pm} +\nu_{e}(\bar{\nu}_e)+ \nu_{\mu}(\bar{\nu_\mu}) $ and $\pi^{0} \rightarrow 2\gamma $. 

For a CSM mass density scaling as $\propto R^{-2}$ \citep[e.g.][]{Chevalier1982}, the number density of the shocked CSM  is written as $ n(R)\approx 4 n_{csm}(R) \simeq 2\times 10^{12}\, R_{\rm in,14}^{-1}\beta_{-1.5}^{-1}(R_{in}/R)^2\,{\rm cm}^{-3}$. Assuming that a fraction $\epsilon_B$ of the post-shock thermal energy density is magnetic, the strength of the magnetic field is estimated by $ B(R)\simeq 46 \, \epsilon_{B,-4}^{1/2}\beta_{-1.5}^{1/2}R_{\rm in,14}^{-1/2}
(R_{\rm in}/R)^{\alpha_B}\,{\rm G}$. Here, we introduced the notation $Q_x=Q/10^x$ in cgs units, $R_{\rm in}$ is the effective inner radius of the CSM, i.e., the shock breakout radius, $\alpha_B=1$ for a wind-like density profile and $\beta=v_{\rm sh}/c$ where $v_{\rm sh}$ is the forward shock velocity.

\section{Numerical framework}
\vspace{-0.1in}
We use the numerical code originally presented in \cite{mastichiadis_kirk1995} which includes:
synchrotron emission, inverse Compton scattering (ICS) on background photons and on synchrotron photons (i.e. synchrotron self-Compton), synchrotron self-absorption (SSA), photon-photon ($\gamma\gamma$) absorption and photohadronic ($p\gamma$) interactions. We expanded the original code by including two physical processes, namely \pp collisions and the adiabatic expansion of the source \citep[for the relativistic equivalent, see][]{petropoulou_mastichiadis2009}. The inclusion of secondary particle production by  \pp collisions was based on \cite{kelner2006}.  

\section{Results}
\vspace{-0.1in}
In Fig.~1 we present snapshots of the MW non-thermal emission of a fiducial SN Type IIn located at $d_L=5$ Mpc (left panel). The radio ($10^3$ GHz) and GeV $\gamma$-ray light curves are illustrated in the right panel. 
The parameters used for this indicative example are summarized below:
\begin{center}
\begin{tabular}{l c c}
 \hline
 Parameter & Symbol & Default value \\
 \hline
 Power-law index for the magnetic field profile & $\alpha_B$ & 1\\
 Magnetic energy density fraction & $\epsilon_B$ & $10^{-4}$\\
 Accelerated proton energy fraction & $\epsilon_p$ & $3\times10^{-3}$\\
 Electron-to-proton ratio & $K_{e,p}$ & 0.01 \\
 Proton injected luminosity (erg s$^{-1}$)  & $L_p$ & $10^{41}$ \\
 Electron injected luminosity(erg s$^{-1}$) & $L_e$ & $10^{39}$ \\
 Power-law slope of the shock-accelerated particles & p & 2 \\
 Maximum Lorentz factor &$\gamma_{p,e,\max}$ & $10^6$\\ 
 Minimum Lorentz factor & $\gamma_{p,e,\min}$ & $10^{0.1}$\\
 Breakout radius (cm) & $R_{in}$&$10^{14}$\\
 Shock velocity at breakout radius & $v_s$&$0.03c$\\
 Electron temperature of unshocked CSM (K) &$T_e$ & $10^5$\\
 \hline
\end{tabular}
\end{center}

\vspace{-0.2in}

%{\bf [Dimitri:] add the parameter values we used for this example. You should also mention Te for free-free absorption, gpmax, gemax, Lp, Le, ep, Kep etc.!}
\begin{figure*}[h]
\centering
\includegraphics[width=.49\linewidth]{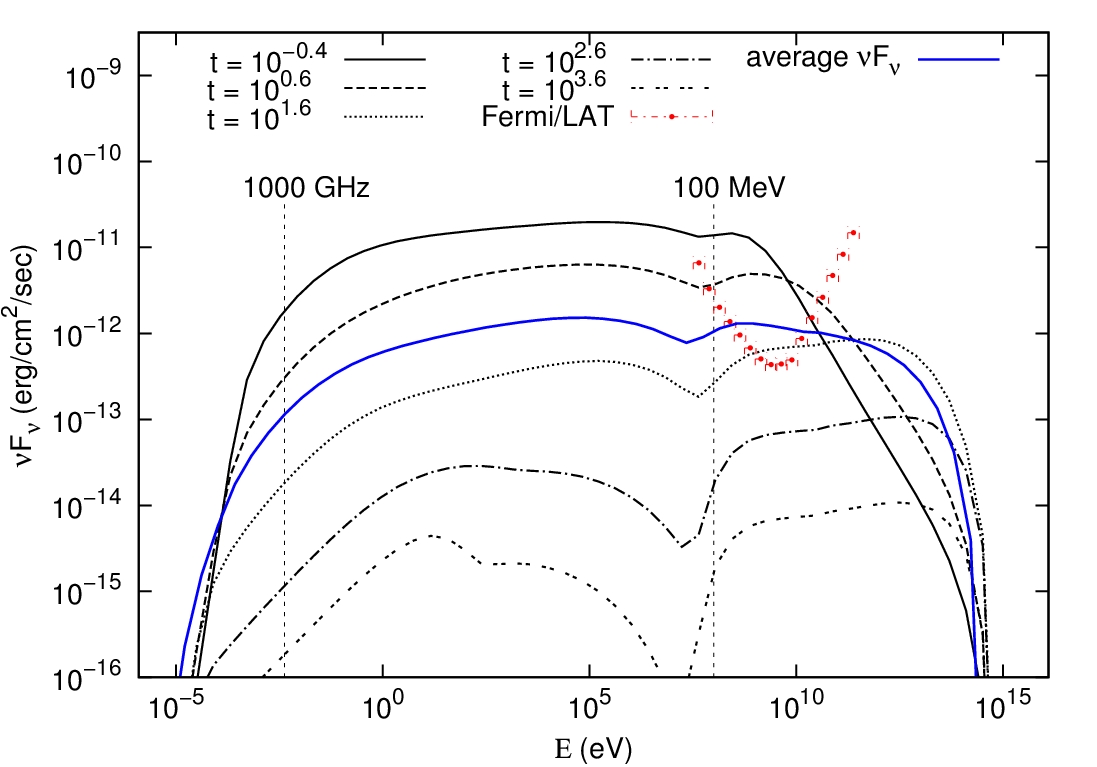}%}
\includegraphics[width=.49\linewidth]{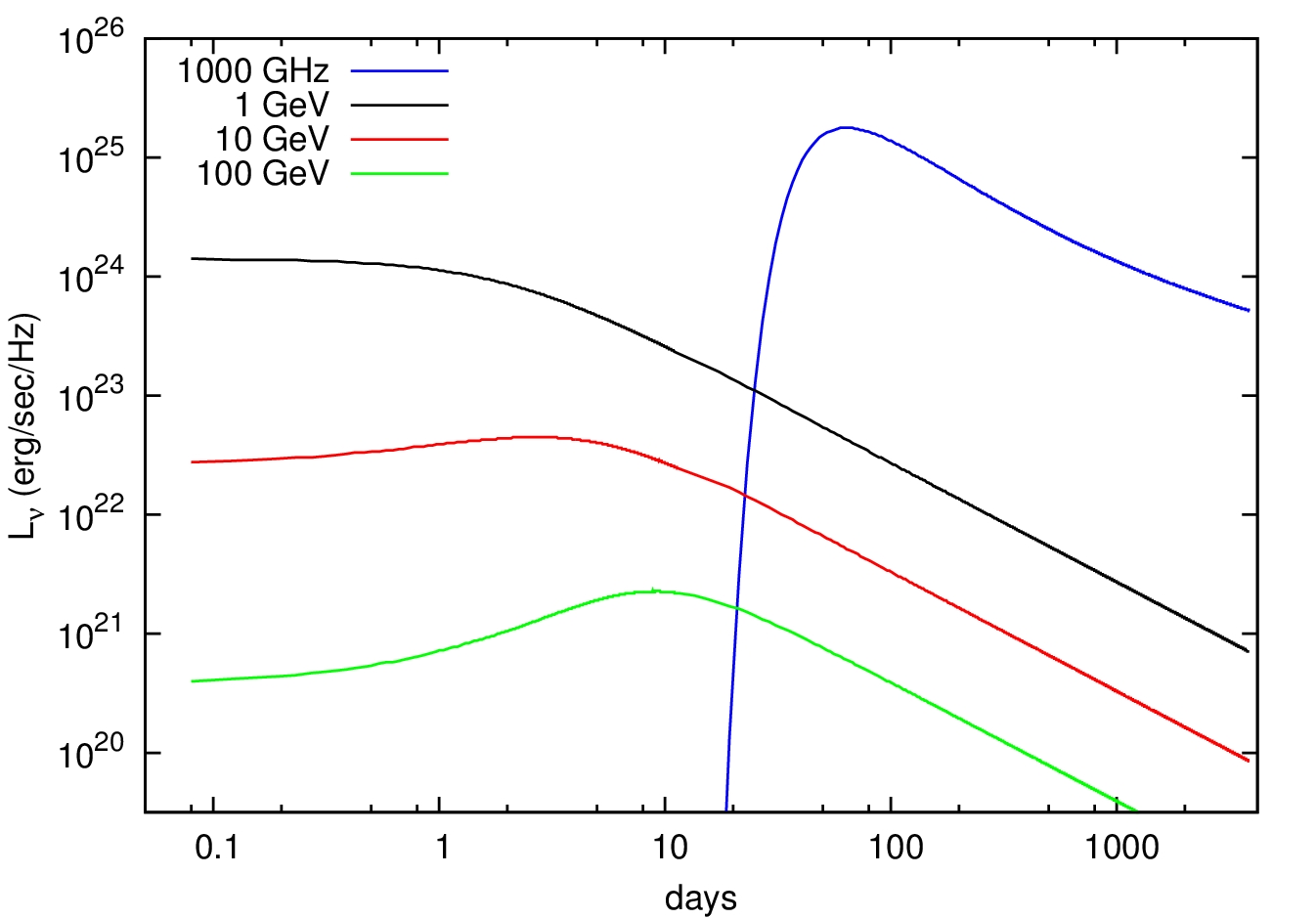}
\caption{{\bf Left panel:} Temporal evolution (for $t\leq 10^{3.6} {\rm d}\simeq 11$yr) of the MW non-thermal spectrum of a fiducial SN Type IIn at $d_L=5$ Mpc. The radio emission is synchrotron self-absorbed at $E\sim 10^{-5}$ eV, while the spectrum is dominated by the synchrotron emission of primary and secondary electrons up to $E\sim10$ MeV. The $\gamma$-ray emission from $\pi^0$ decays extends from $\sim 100$ MeV up to $E\sim 0.1$ PeV for protons accelerated to $\sim 1$ PeV. The average MW spectrum predicted by the model for a time interval of 10 yr is overplotted with a blue line. Here, the radio emission does not include free-free absorption. The \emph{Fermi}-LAT point-source sensitivity for a 10-yr exposure is depicted with red symbols ({\sl \href{http://www.slac.stanford.edu/exp/glast/groups/canda/lat\_Performance.htm}{http://www.slac.stanford.edu/exp/glast/groups/canda/lat\_Performance.htm}}). {\bf Right panel:} Monochromatic light curves at $10^3$~GHz, 1 GeV, 10~GeV and 100~GeV. The radio light 
curve is calculated after taking into account the free-free absorption by thermal electrons of the unshocked CSM with $T_e=10^5~K$. For display reasons, the GeV light curves are shifted upwards by a factor of 7 in logarithm. The early-time flattening of the $\gamma$-ray light curves is due to the $\gamma \gamma$ absorption on the non-thermal synchrotron photons. The $10^3$ GHz emission peaks only after $\sim$10 d due to the early-time free-free absorption. At later times the luminosity decays as a power-law, i.e., $L_{\nu}\propto t^{-0.7}$ at $10^3$ GHz and $L_{\nu}\propto t^{-1}~$ at $>1$ GeV. } 
\end{figure*}

\section{Conclusions and outlook}
\vspace{-0.1in}
We presented the temporal evolution of the non-thermal MW emission from interaction-powered SNe (e.g. Type IIn) while focusing on the $\gamma$-ray emission produced by \pp collisions between the shock-accelerated protons and the non-relativistic protons of the CSM. We have shown that for sufficiently dense CSM
($n_{csm} > 10^{11}$ cm$^{-3}$) or for a nearby SN Type IIn ($d_L<10$Mpc), the $\gamma$-ray  ($>100$MeV) emission would be detectable by \emph{Fermi}-LAT. The non-detection of $\gamma$-rays from extragalactic interaction-powered SNe by the \emph{Fermi}-LAT so far, can be thus used to constrain the CR acceleration efficiency.  We have explicitly shown that the $\gamma$-ray emission ($\sim 100$~GeV) would be attenuated at early times (i.e. $<10$~d) due to internal  $\gamma \gamma$  absorption. The optical SN emission, which typically peaks several ($\gtrsim 10$ d) days after the shock break out, may also attenuate the $\sim$100 GeV emission. We plan to include this in our numerical code by modelling the SN optical radiation as an external photon field of variable photon number density. Another significant process for producing $\gamma$-rays in such dense environments is the bremsstrahlung radiation of relativistic electrons. Although the results presented here do not include relativistic bremsstrahlung radiation,  we plan to implement it in the code both as an energy loss process for electrons and a photon production process. We have also verified that the effect of $p\gamma$ interactions upon the synchrotron photons is negligible. Yet, $p\gamma$ interactions with the X-ray bremsstrahlung photons produced by the hot shocked plasma 
may be important for creating additional secondaries, especially at early times when the X-ray luminosity is high and the source is compact. 

\vspace{-0.2in}

% Do not delete the next line
\small  % Do not delete
%
%%% Comment the following line if you do not have acknowledgments.
\section*{Acknowledgments}   % Do not delete if you declare acknowledgments
%%% ACKNOWLEDGMENTS
M.P. acknowledges  support  for  this  work by NASA through Einstein Postdoctoral Fellowship grant number PF3 140113 awarded by the Chandra X-ray Center, which is operated  by  the  Smithsonian  Astrophysical  Observatory  for  NASA under contract NAS8-03060.

%%% BIBLIOGRAPHY
\bibliographystyle{aj}
\small
\bibliography{proceedings}

\end{document}